\documentclass{article}

\usepackage[english]{babel}

\usepackage[letterpaper,top=2cm,bottom=2cm,left=3cm,right=3cm,marginparwidth=1.75cm]{geometry}

\usepackage{amsmath}
\usepackage{graphicx}
\usepackage{booktabs} 
\usepackage[colorlinks=true, allcolors=blue]{hyperref}
\usepackage{accents}
\usepackage{cite}
\newcommand{\ubar}[1]{\underaccent{\bar}{#1}}

\title{Quantum Properties and Gravitational Field of a Proper Time Oscillator}
\author{Hou Y. Yau \footnote{San Francisco State University, 1600 Holloway Avenue, CA, USA}\footnote{Email: hyau@mail.sfsu.edu}}

\begin{document}
\maketitle

\abstract{We find that a field with oscillations of matter in proper time has the properties of a zero-spin bosonic field. A particle observed in this field is a proper time oscillator. Neglecting all quantum effects, a proper time oscillator can mimic a point mass at rest in general relativity. The spacetime outside a 'stationary' proper time oscillator is a Schwarzschild field.}

\maketitle

\section{Introduction}
\label{intro}

A quantum field is defined as a sum of creation and annihilation operators with formulations akin to those developed for a quantum harmonic oscillator. However, the oscillators created/annihilated in a quantum field are treated as coupled particles versus the quantized energy levels of a quantum harmonic oscillator. Although the natures of the two oscillators are different, they have many shared features in their formulations. Because of these analogies, can there be a hidden symmetry relating to the two kinds of oscillators? On the other hand, when we study a quantum harmonic oscillator, oscillations are considered only in the spatial frame. However, as stipulated by relativity, space and time are supposed to be treated on an equal footing. If so, can matter also oscillate in time? Can this temporal oscillation be a missing link explaining the analogies between a quantum field's formulations and a quantum harmonic oscillator? 

In response to the lack of symmetry between time and space in the formulations of quantum theory, we assume matter can oscillate in proper time \cite{Yau18,Yau19,Yau16,Yau20,Yau21,Yau17,Yau20a,Yau12}. We find that a field with oscillations of matter in proper time has the properties of a zero-spin bosonic field \cite{Yau18,Yau19,Yau16}. The proper time oscillations in the matter field are quantized. The particles created/annihilated are oscillators in proper time. 

Interestingly, the internal time in this matter field can be reckoned as a self-adjoint operator \cite{Yau20}. This outcome has no contradiction with the conundrum of Pauli's theorem \cite{Pauli80,Srinivas81} since the internal time and the Hamiltonian of this system are not a conjugate pair. This proposition is different from the results attained from other approaches that also treat time as a dynamic variable \cite{Muga00,Egusquiza00,Galapon02,Wang07,Muga08,Muga09,Galapon09,Lee83,Lee87,Greenberger10,Bauer14,Aharonov98,Olkhovsky07,Ordonez09,Kiukas09,Brunetti10,Hegerfeldt10,Yearsley11,Strauss11,Arsenovic12,Madrid13,Kullie18,Ashmead19}.  

Following the same concepts developed, we can obtain an uncertainty relation for the internal time of a field with proper time oscillators \cite{Yau21}. This uncertainty relation is similar to the one for a quantum harmonic oscillator, except the oscillation is in time and not space. Contrary to most of the time-energy uncertainty relation contemplated so far \cite{Bohm61,Holevo82,Brunetti02,Ozawa04,Busch90A,Busch90B,Hilgvoord06,busch07}, the internal proper time and its conjugate are self-adjoint operators. These results provide an explanation why the formulations of a quantum field have so many similarities with a quantum harmonic oscillator. 

We can apply the concepts of proper time oscillation outside quantum theory. If we neglect all quantum effects, a proper time oscillator can be treated as a 'stationary' classical object, equivalent to a point mass at rest in general relativity. Under this assumption, we expect the proper time oscillator to curve the surrounding spacetime and generate a gravitational field; its solution shall be the Schwarzschild metric. These results have been recently affirmed in refs. \cite{Yau17,Yau20a}. The self-adjoint time operator, proper time uncertainty relation, and generation of a Schwarzschild field are properties that could reduce some of the differences between quantum theory and general relativity. 

The paper is organized as follows: Section 2 introduces the concepts of proper time oscillation. Sections 3 and 4 provide an overview of a proper time oscillator's quantum properties and gravitational field. Section 5 further elaborates on the properties of a proper time oscillator and clarifies certain concepts. The last section is reserved for discussions and conclusions.

\section{Internal Time and Proper Time Oscillation}
\label{sec:2}

What do we mean by oscillation in proper time? To illustrate the idea, let us consider an analogy with a particle traveling at an average velocity $\mathbf{v}$. The particle also oscillates with an angular frequency $\omega$ and an amplitude $\mathring{\mathbf{X}}$, i.e.,
\begin{equation}\label{eq:particlex}
\mathring{\mathbf{x}}_f = \mathbf{v}t-\mathring{\mathbf{X}}\sin(\omega t),
\end{equation}
where $\mathring{\mathbf{x}}_f$ is the position of the particle. To a stationary observer, the particle has a varying velocity. Suppose the angular frequency $\omega$ is large, and the amplitude $\mathring{\mathbf{X}}$ is small; the particle will appear to travel with a constant velocity if the instrument used by the observer is not sensitive enough to detect the slight variation of the oscillation. The properties of this model can be readily derived from classical mechanics. In this paper, we will investigate a similar model but replace the motions in space with the propagation and oscillation of a particle in proper time.

Consider a Minkowski coordinate system $(t,\mathbf{x})$. The coordinate time $t$ is measured by the clock of a stationary observer $O$ at spatial infinity. From what we have learned from relativity, a particle travels along a time-like geodesic. Instead of propagating smoothly, let us assume the proper time of a stationary particle also oscillates, i.e.,
 \begin{equation}\label{eq:propparticlet}
\mathring{t}_f = t+\mathring{t}_{d} = t-\mathring{T}_0\sin(\omega_0 t),
\end{equation}
where
\begin{equation}\label{eq:t_d}
\mathring{t}_{d} = -\mathring{T}_0\sin(\omega_0 t).
\end{equation}
Note that Eq. (\ref{eq:propparticlet}) is an analogy of Eq. (\ref{eq:particlex}), except motion in space is replaced by propagation in time.

The 'internal time' $\mathring{t}_f$ is an assumed intrinsic property of a particle, which oscillates with an amplitude $\mathring{T}_0$ and an angular frequency $\omega_0$. Without the assumed oscillation, the defined internal time is just the coordinate time, i.e., $\mathring{t}_f=t$. The amplitude $\mathring{T}_0$ is analogous to the amplitude of a classical oscillator, except the oscillation is in time and not in space. As noted, the coordinate time $t$ remains a parameter. Only the temporal oscillation displacement $\mathring{t}_d$ is the dynamic component of the internal time $\mathring{t}_f$. From Eq. (\ref{eq:propparticlet}), the rate of the internal time relative to the coordinate time is,
\begin{equation} \label{eq:R+t}
\frac{\partial \mathring{t}_f}{\partial t}=1-\mathring{T}_0 \omega_0\cos(\omega_0t),
\end{equation} 
which has an average rate of 1.

The 'flowing' coordinate time $t$ labels the equilibrium position of the proper time oscillation. The temporal oscillation displacement $\mathring{t}_{d}$ is measured against this 'equilibrium'. If the clock at spatial infinity $O$ is not sensitive enough to pick up the internal time oscillation, the particle will appear to travel along a faux
time-like geodesic as if there is no oscillation. On the other hand, the proper time oscillator is stationary at a fixed coordinate $\mathbf{x}$ of the spatial frame. A particle at rest has no spatial oscillation displacement, i.e., $\mathring{\mathbf{x}}_d=0$. 

We can extend the idea of proper time oscillation to a plane wave $\zeta_{to}$. Let us assume all the matters inside a plane wave have oscillations in proper time. We define the internal time of matters as,
\begin{equation} \label{eq:proptfluct}
t_{f}(t,\mathbf{x})=t+\textrm{Re}[\zeta_{t0}(t,\mathbf{x})]=t-T_0\sin(\omega_0 t),
\end{equation} 
where
\begin{equation} \label{eq:proptfluct}
\zeta_{t0}(t,\mathbf{x})=-iT_0e^{-i\omega_0t}.
\end{equation}
The temporal oscillation displacement is,
\begin{equation} \label{eq:t_dplane wave}
t_{d}(t,\mathbf{x})=\textrm{Re}[\zeta _{t0}(t,\mathbf{x})]=-T_0\sin(\omega_0 t).
\end{equation} 
The rate of the internal time relative to the coordinate time is,
\begin{equation} \label{eq:tfluct'}
\frac{\partial t_{f}(t,\mathbf{x})}{\partial t}=1-\omega_0 T_0\cos(\omega_0 t).
\end{equation}
Again, this internal time rate has an average of 1. In addition, all the matters inside the plane wave are stationary in the spatial frame. These matters at rest have no spatial oscillation displacement, i.e., ${\mathbf{x}}_d(t,\mathbf{x})=0$. The properties of this plane wave are analogous to those for the proper time oscillator, except we apply proper time oscillations to all matters inside a plane wave.

Under a Lorentz transformation to another reference frame $O'$, matter in the plane wave will appear to oscillate in both space and time, i.e.,
\begin{equation} \label{eq:tfluct}
t'_f=t'+t'_d=t'+\textrm{Re}(\zeta_{t\mathbf{k}})=t'+T_\mathbf{k}\sin(\mathbf{k}\cdot\mathbf{x}'-\omega t'),
\end{equation}
\begin{equation} \label{eq:xfluct}
\mathbf{x}'_f=\mathbf{x}'+{\mathbf{x}}'_d=\mathbf{x}'+\textrm{Re}(\zeta_{\mathbf{x}\mathbf{k}})=\mathbf{x}'+\mathbf{X}_\mathbf{k}\sin(\mathbf{k}\cdot\mathbf{x}'-\omega t'),
\end{equation}
where
\begin{equation} \label{eq:oscillations}
 t'_d=\textrm{Re}(\zeta_{t\mathbf{k}}), \quad
{\mathbf{x}}'_d= \textrm{Re}(\zeta_{\mathbf{x}\mathbf{k}}),   
\end{equation}
\begin{equation} \label{eq:zetat}
\zeta_{t\mathbf{k}}=-iT_\mathbf{k}e^{i(\mathbf{k}\cdot\mathbf{x}'-\omega t')}, 
\end{equation}
\begin{equation} \label{eq:zetax}
\zeta_{\mathbf{x}\mathbf{k}}=-i\mathbf{X}_\mathbf{k}e^{i(\mathbf{k}\cdot\mathbf{x}'-\omega t')},    
\end{equation}
\begin{equation} \label{eq:TX}
T_\mathbf{k}=(\omega/\omega_0)T_{0}, \quad \mathbf{X}_\mathbf{k}=(\mathbf{k}/\omega_0)T_{0}.
\end{equation}
We assume frame $O$ travels at a velocity $\mathbf{v}=\mathbf{k}/\omega$ relative to frame $O'$. Apart from the oscillations, matter in this plane wave travels with an average velocity $\mathbf{v}$. 

As we shall note, $(T_\mathbf{k},\mathbf{X}_\mathbf{k})$ is a 4-amplitude Lorentz transformation of $(T_0,\mathbf{0})$, i.e.,
\begin{equation} \label{eq:amplitudes}
 \mid T_\mathbf{k}\mid ^2=\mid T_{0} \mid^2+\mid \mathbf{X}_\mathbf{k}\mid ^2,  
\end{equation}
and the natural units $c=\hbar=1$ are adopted in this paper. The derivative of the internal time relative to the coordinate time is
\begin{equation} \label{eq:tfluct'}
\frac{\partial t'_{f}}{\partial t'}=1-\omega T_\mathbf{k}\cos(\mathbf{k}\cdot\mathbf{x}'-\omega t').
\end{equation}

In the above analysis, we adopted the Lagrangian wave mechanics formulation. The temporal and spatial displacements $t'_d(t',\mathbf{x}')$ and $\mathbf{x}'_d(t',\mathbf{x}')$ are measured against the undisturbed state $(t',\mathbf{x}')$. In the Lagrangian formulation, $\mathbf{x} '_d(t',\mathbf{x}')$ tells us the spatial displacement of matter from the undisturbed coordinate $\mathbf{x}'$ at time $t'$. Similarly, $ t'_d (t',\mathbf{x}')$ is the difference between the matter's internal time and the coordinate time $t'$. Matter originally at $\mathbf{x}'$ will be displaced to $\mathbf{x}'_f=\mathbf{x}'+\mathbf{x}'_d$, and measures a time $t'_f=t'+t'_d$. As shown, matter in this plane wave has vibrations in both time and space. Note that $\zeta_{t\mathbf{k}}$ and $\zeta_{\mathbf{x}\mathbf{k}}$ form a Lorentz covariant plane wave with a 4-vector amplitude, 
\begin{equation} \label{eq:array2}
\left[
  \begin{array}{ c }
     \zeta_{t\mathbf{k}} \\
     \zeta_{\mathbf{x}\mathbf{k}}
  \end{array} \right]
=  -i   
\left[
  \begin{array}{ c }
     T_\mathbf{k} \\
     \mathbf{X}_\mathbf{k}
  \end{array} \right]e^{i(\mathbf{k}\cdot\mathbf{x}'-\omega t')}. 
\end{equation}

\section{Matter Field with Proper Time Oscillations}
\label{sec:3}

We have defined internal time and proper time oscillation in the previous section. Our next task is to investigate what properties the temporal oscillation can produce and compare them with those derived from quantum theory and general relativity. Summarizing the findings from our previous articles \cite{Yau18,Yau19,Yau16,Yau20,Yau21}, we have obtained the following results:

1. Proper time oscillation in a matter field is quantized \cite{Yau18,Yau19,Yau16}.

2. A matter field with proper time oscillations has the properties of a bosonic field. The particles observed are proper time oscillators \cite{Yau18,Yau19,Yau16}. 

3. The internal time observed in a matter field can be treated as a self-adjoint operator \cite{Yau20}.

4. A proper time uncertainty relation can be derived analogously to the one for a quantum harmonic oscillator, except the oscillation is in time and not in space \cite{Yau21}.

\subsection {Quantization of Proper Time Oscillation \cite{Yau18,Yau19,Yau16}} \label{sec:3.1}

Let us define a plane wave, 
\begin{equation} \label{eq:zeta}
\zeta_\mathbf{k}=\frac{T_{0\mathbf{k}}}{\omega _0}e^{i(\mathbf{k}\cdot\mathbf{x}-\omega t)}.
\end{equation}
The temporal and spatial oscillation displacements\footnote{For convenience purposes, we have dropped the prime symbol from the coordinate $(t',\mathbf{x}')$ in the rest of the paper.}  from Eqs. (\ref{eq:zetat}) and (\ref{eq:zetax}) can be written in terms of $\zeta_\mathbf{k}$, i.e.
\begin{equation} \label{eq:zetat1}
\zeta_{t\mathbf{k}}=\partial_0\zeta_\mathbf{k}=-iT_\mathbf{k}e^{i(\mathbf{k}\cdot\mathbf{x}-\omega t)}, 
\end{equation}
\begin{equation} \label{eq:zetax1}
\zeta_{\mathbf{x}\mathbf{k}}=-\mathbf{\nabla}\zeta_\mathbf{k}=-i\mathbf{X}_\mathbf{k}e^{i(\mathbf{k}\cdot\mathbf{x}-\omega t)}.    
\end{equation}
Therefore, $\zeta_{\mathbf{k}}$ is a plane wave with both temporal and spatial oscillations. To obtain the oscillations, we simply takes the derivatives of $\zeta_{\mathbf{k}}$ as shown in Eqs. (\ref{eq:zetat1}) and (\ref{eq:zetax1}). Also, $\zeta_{\mathbf{k}}$ serves another purpose; it represents a plane wave with matter moving at an average velocity $\mathbf{v}$.

The plane wave $\zeta_{\mathbf{k}}$ and its conjugate $\zeta_{\mathbf{k}}^\ast$ satisfy the Klein Gordon equation:
\begin{equation} 
\label{eq:kg}
\partial_u\partial^u\zeta_{\mathbf{k}} +\omega_0^2\zeta_{\mathbf{k}}=0,
\end{equation}
\begin{equation} 
\label{eq:kgc}
\partial_u\partial^u\zeta_{\mathbf{k}}^\ast+\omega_0^2\zeta_{\mathbf{k}}^\ast=0.
\end{equation}
For a system that can have multiple numbers of particles with mass m, the Lagrangian density and Hamiltonian density of $\zeta_{\mathbf{k}}$ are defined as,
\begin{equation} \label{eq:lagrangian}
\mathcal{L}_{\mathbf{k}}=\frac{m \omega_0^2}{2V}[(\partial^u\zeta_{\mathbf{k}}^\ast)(\partial_u\zeta_{\mathbf{k}})-\omega_0^2\zeta_{\mathbf{k}}^\ast\zeta_{\mathbf{k}}],
\end{equation}
\begin{equation} \label{eq:Hamiltonian}
\mathcal{H}_{\mathbf{k}}=\frac{m \omega_0^2}{2V}[(\partial_0\zeta_{\mathbf{k}}^\ast)(\partial_0\zeta_{\mathbf{k}})+(\mathbf{\nabla}\zeta_{\mathbf{k}}^\ast)\cdot(\mathbf{\nabla}\zeta_{\mathbf{k}})+\omega_0^2\zeta_{\mathbf{k}}^\ast\zeta_{\mathbf{k}}].
\end{equation}
Volume $V$ is a cube with periodic boundary conditions imposed on the walls. Moreover, we have adopted de Broglie's angular frequency as the mass-energy of the particles, $\omega_0=m$ \cite{Broglie24}.

The proper time oscillation is quantized. To understand the reason, let us consider a plane wave that has oscillation of matter in proper time only ($\vert \mathbf{k}\vert=0$, and $\vert \mathbf{X}\vert=0$), i.e.,
\begin{equation} \label{eq:zeta0}
\zeta_{0}=\frac{T_{0}}{\omega _0}e^{-i\omega_0 t}.
\end{equation}
The Hamiltonian density from Eq. (\ref{eq:Hamiltonian}) is,
\begin{equation} \label{eq:properham}
\mathcal{H}_0=(\frac{m\omega_0^2}{V}) T_0^\ast T_0.
\end{equation}
As we shall note, matter in this plane wave has no motion and oscillation in the spatial frame. Therefore, the energy generated by the proper time oscillations shall belong to some intrinsic energy of the system. However, the field we are considering is 'free' with no charges or force fields. The only energy present is the intrinsic mass-energy of the matter inside the plane wave. For this reason, we adopt the energy in $\mathcal{H}_0$ as the intrinsic mass-energy of matter inside volume $V$.

For a plane wave with $n$ number of particles, the Hamiltonian density shall be $\mathcal{H}_0=nm/V$. Compare with Eq. (\ref{eq:properham}), the energy $E$ in volume $V$ is,
\begin{equation} \label{eq:quantized m}
E=nm=m\omega _0^2T_0^\ast T_0,
\end{equation}
which leads to a quantization condition,
\begin{equation} \label{eq:quantized T0}
\omega _0^2T_0^\ast T_0=n.
\end{equation}
Since the number of particles is discrete, the proper time oscillation shall be quantized (or more strictly speaking, it is $T_0^\ast T_0$ that is quantized). We shall treat a matter field with proper time oscillations as a quantized field. 

\subsection {Bosonic Field and Proper Time Oscillator \cite{Yau18,Yau19,Yau16}}

We can obtain a real scalar field by superposition of the plane waves ${{\zeta}}_{\mathbf{k}}$ and their conjugates ${{\zeta}}^\ast_{\mathbf{k}}$, i.e.,
\begin{equation} \label{eq:zetasuper}
\zeta({x}) = \sqrt{\frac{\omega_0}{\omega}}\sum\limits_{\mathbf{k}} \frac{[{{\zeta}}_{\mathbf{k}}({x}) + {{\zeta}}^\ast_{\mathbf{k}}({x})]}{\sqrt{2}} \\
= \sum\limits_{\mathbf{k}}(2\omega \omega_0)^{-1/2}  [T_{0\mathbf{k}} e^{-i{k} {x}} + T^\ast_{0\mathbf{k}} e^{i{k}{x}}],
\end{equation}
where $\sqrt{{\omega_0}/{\omega}}$ is a normalization factor. Following the same concepts developed in quantum theory, we can transform a classical field into a quantized field through canonical quantization. In other words, quantities such as $\zeta(x)$, $T_{0\mathbf{k}}$ and $T_{0\mathbf{k}}^\dagger$ are promoted to operators. As a quantized field, we can relate $\zeta({x})$ with the bosonic field $\varphi({x})$ in quantum theory, i.e.
\begin{equation} \label{eq:varphi2}
\varphi({x})=\zeta({x})\sqrt{\frac{\omega_0^3}{V}}= \sum\limits_{\mathbf{k}}(2 \omega V)^{-1/2} [a_\mathbf{k} e^{-i{k} {x}} + a^\dagger_\mathbf{k} e^{i{k} {x}}],
\end{equation}
where
\begin{equation} \label{eq:a}
a_\mathbf{k}=\omega_0 T_{0\mathbf{k}}, \quad a^\dagger_\mathbf{k}=\omega_0 T^\dagger_{0\mathbf{k}},
\end{equation}
are the annihilation and creation operators. 

As shown, a field with proper time oscillations has the properties of a bosonic field. The properties of $\zeta({x})$ can be derived from the standard quantum theory by replacing the proper time amplitudes and their adjoints with the annihilation and creation operators through Eq. (\ref{eq:a}). (For example, Eq. (\ref{eq:quantized T0}) derived from the proper time oscillations becomes the number operator in a quantum field.) Introducing proper time oscillations to a matter field leads us to the same properties of a bosonic field derived from the quantum field theory. 

As shown in Eq. (\ref{eq:a}), the operator $T^\dagger_{0}$ creates a particle with proper time oscillation. This proper time oscillator is the same one we have defined in Section 2. Under the quantization condition Eq. (\ref{eq:quantized T0}), a particle with mass $m$ has a unique amplitude,
\begin{equation} \label{eq:T0}
\vert\mathring{T}_0\vert =1/\omega_0.    
\end{equation}
From Eq. (\ref{eq:propparticlet}), the particle's internal time is,
\begin{equation}\label{eq:propparticlet1}
\mathring{t}_{f}=t+\mathring{t}_{d}=t-\frac{\sin(\omega_0 t)}{\omega_0}.
\end{equation}
Its time rate is,
\begin{equation} \label{eq:timerate}
\frac{\partial \mathring{t}_{f}}{\partial t}=1-\cos(\omega_0t),
\end{equation}
which is bounded between 0 and 2. As a result, the internal time $\mathring{t}_f$ bounces along the time-like geodesic but never goes back to its past. Since the de Broglie's angular frequency of a particle (e.g., $\omega _0=7.6\times10^{20 }s^{-1}$ for electron) is so rapid, it is challenging to detect its effects. If the experiment is not sensitive enough to pick up the oscillation, the particle will appear to propagate along a faux time-like geodesic. 

\subsection{Self-Adjoint Internal Time Operator \cite{Yau20}} \label{sec3.3}

Based on Eq. (\ref{eq:lagrangian}) and promote the Lagrangian as an operator, the conjugate momenta of $\zeta({x})$ is,
\begin{equation} \label{eq:eta}
\eta({x})=\frac{\partial \mathcal{L}}{\partial [\partial_0\zeta ({x})]}=\frac{-i\omega_0^3}{\sqrt2 V} \sum\limits_{\mathbf{k}}  [\tilde{T}_\mathbf{k} e^{-i{k}{x}}-\tilde{T}^\dagger_\mathbf{k} e^{i{k} {x}}],
\end{equation}
where $\tilde{T}_{\mathbf{k}}=T_{0\mathbf{k}} \sqrt{{\omega }/{\omega_0}}$. We can show that $\zeta(x)$ and $\eta(x)$ satisfy the equal-time commutation relations,
\begin{equation} \label{zeta-eta}
[\zeta (t,\mathbf{x}),\eta (t,\mathbf{x}')]=i \delta (\mathbf{x}-\mathbf{x}'),
\end{equation}
\begin{equation} \label{zeta-eta1} 
[\zeta (t,\mathbf{x}),\zeta (t,\mathbf{x}')]=[\eta (t,\mathbf{x}),\eta (t,\mathbf{x}')]=0.
\end{equation}
Applying Eq. (\ref{eq:zetat1}), the displaced time in the matter field is linearly related to $\eta({x})$, i.e.,
\begin{equation} \label{eq:zetat(x)}
t_d({x})=\zeta_t({x})=\partial_0\zeta({x})= \sum\limits_{\mathbf{k}} \frac{-i}{\sqrt2}  [\tilde{T}_\mathbf{k} e^{-i{k} {x}}-\tilde{T}^\dagger_\mathbf{k} e^{i{k}{x}}]=\frac{\eta({x})V}{\omega_0^3}.
\end{equation}
Therefore, $t_d(x)$ and $\zeta(x)$ also form a conjugate pair and satisfy similar equal-time commutation relations, 
\begin{equation} \label{zeta-zetat}(\frac{\omega_0^3}{V})[\zeta (t,\mathbf{x}),t_d (t,\mathbf{x}')]=  i\delta (\mathbf{x}-\mathbf{x}'),
\end{equation}
\begin{equation} \label{zeta-zetat1}
[t_d (t,\mathbf{x}),t_d (t,\mathbf{x}')]=0.
\end{equation}

In a real scalar quantum field, $\varphi({x})$ and its conjugate momenta $\Pi(x)$ are self-adjoint operators. Because $\varphi({x})$ and $\zeta({x})$ are linearly related as shown in Eq. (\ref{eq:varphi2}), the quantities $\zeta({x})$, $\eta({x})$ and $t_d({x})$ must also be self-adjoint operators. This outcome is what we expect since the displaced time $t_d$ oscillates relative to the external time $t$. Its spectrum is not bounded. 

The internal time in a matter field is,
\begin{equation} \label{eq:tfluctfield}
t_f(t,\mathbf{x})=t+t_d(t,\mathbf{x}).
\end{equation}
As discussed, $t$ is a parameter, but $t_d(t,\mathbf{x})$ is a self-adjoint operator. These two time variables are treated differently in theory. The only dynamic component of the internal time is the temporal displacement $t_d(t,\mathbf{x})$. Being the sum of $t$ and $t_d(t,\mathbf{x})$, the internal time $t_f$ must also be a self-adjoint operator. We can establish equal-time commutation relations between $\zeta (t,\mathbf{x})$ and $t_f(t,\mathbf{x})$ based on the results from Eqs. (\ref{zeta-zetat}) and (\ref{zeta-zetat1}),
\begin{equation} \label{zeta-zetat2}
(\frac{\omega_0^3}{V})[\zeta (t,\mathbf{x}),t_f (t,\mathbf{x}')]= i\delta (\mathbf{x}-\mathbf{x}'),
\end{equation}
\begin{equation} 
[t_f (t,\mathbf{x}),t_f (t,\mathbf{x}')]=0.
\end{equation}  
As we will explain in Section \ref{sec5.3}, the internal time $t_f$ can be treated as a self-adjoint operator without contradicting Pauli's theorem \cite{Pauli80,Srinivas81}. 

\subsection{Proper Time Oscillator Uncertainty Relation \cite{Yau21}} \label{Sect3.4}

Let us consider a real scalar field that has oscillations of matter in proper time only, i.e.,
\begin{equation} \label{eq:zetasuper0}
\zeta' = \frac{1}{\sqrt{2}}[{\zeta}_0 + {\zeta}_0^\dagger]=\frac{1}{\sqrt2\omega_0}  [T_0 e^{-i{\omega_0} t}+T_0^\dagger e^{i{\omega_0}t}]. 
\end{equation}
The particles observed are motionless in the spatial frame. Applying Eq. (\ref{eq:zetat1}), the displaced time $t'_d$ and the displaced time rate $u'_d$ are,
\begin{equation} \label{eq:zetat0}
t'_d=\zeta'_{t}=\partial_0\zeta'=\frac{-i}{\sqrt2}  [T_0 e^{-i{\omega_0} t}-T_0^\dagger e^{i{\omega_0}t}]=\frac{-i}{\sqrt2\omega_0}  [a e^{-i{\omega_0} t}-a^\dagger e^{i{\omega_0}t}],
\end{equation}
\begin{equation} \label{eq:zetat0'}
u'_d=\partial_0t'_d=\frac{-\omega_0}{\sqrt2}  [T_0 e^{-i{\omega_0} t}+T_0^\dagger e^{i{\omega_0}t}]=\frac{-1}{\sqrt2}  [a e^{-i{\omega_0} t}+a^\dagger e^{i{\omega_0}t}].
\end{equation}
Based on Eq. (\ref{eq:Hamiltonian}), the Hamiltonian density of the proper time real scalar field can be written as,
\begin{equation} \label{eq:Hamiltont1}
H'=\frac{1}{2} (m\omega_0^2 {t'_d}^2 + \frac{{P'_d}^2}{m})=\omega_0 (a^\dagger a+\frac{1}{2}), 
\end{equation}
where
\begin{equation} \label{eq:Pd}
P'_d=mu'_d.
\end{equation}
This result is analogous to the Hamiltonian density of a quantum harmonic oscillator, except the spatial oscillation is replaced by the temporal oscillation. The two terms after the first equal sign in Eq. (\ref{eq:Hamiltont1}) resemble the `potential' and `kinetic' energy of a classical harmonic oscillator.  

The quantity $ P'_d$ is taking the role of a 'momentum'. The displaced time $ t'_d$ and the 'temporal momentum' $ P'_d$ are analogies of a quantum harmonic oscillator's position and momentum operators. However, we stress that this $ P'_d$ is not the 0-component of the 4-momentum; it is different from the energy of the field.

Based on Eqs. (\ref{eq:zetat0}), (\ref{eq:zetat0'}) and (\ref{eq:Pd}), the displaced time $t'_d$ and the temporal momentum $P'_d$ satisfy an uncertainty relation,  
\begin{equation} \label{eq:deltatp}
\Delta t'_d\Delta P'_d=n+{\frac {1}{2}}\geq \frac {1}{2}.
\end{equation}
This uncertainty relation is obtained from the displaced time and temporal momentum variances. Compare Eq. (\ref{eq:deltatp}) with $\Delta x\Delta p\geq \frac {1}{2}$, the proper time field satisfies an uncertainty relation that is similar to the one obtained for a quantum harmonic oscillator. If a particle has proper time oscillation, it will explain why the formulations of a bosonic field share many similarities with a quantum harmonic oscillator. The introduction of the proper time oscillations allows a more symmetrical treatment between time and space in a matter field.

\section{Schwarzschild Spacetime \cite{Yau17,Yau20a}}

In the following analysis, we will neglect all quantum effects and treat the proper time oscillator as a 'stationary' classical object. Eq. (\ref{eq:propparticlet1}) describes the proper time oscillation, which we will idealize here to be stationary at the spatial origin $\mathbf{x}_0$ of a coordinate system. In the presence of the oscillator, time observed in the spacetime continuum at $\mathbf{x}_0$ will oscillate.

As a part of the spacetime geometry, the proper time oscillation at $\mathbf{x}_0$ has geometrical properties that differ from those at spatial infinity, which is asymptotically flat with no oscillations. The difference in spacetime geometry at two spatially far apart locations implies that the spacetime between cannot be flat. A proper time oscillator can curve its surrounding spacetime and generates a gravitational field. Treating the proper time oscillator as a stationary classical object, it shall have effects equivalent to those for a point mass at rest in general relativity, which we want to demonstrate in this section.

The proper time oscillation at $\mathbf{x}_0$ is a pulse that can be decomposed into a series of plane waves. For a relativistic theory, we shall utilize Lorentz covariant plane waves for the decomposition, i.e.,
\begin{equation} \label{eq:array3}
\left[
  \begin{array}{ c }
     \bar{\xi}_{t\mathbf{k}} \\
     \bar{\xi}_{\mathbf{x}\mathbf{k}}
  \end{array} \right]
=  -i   
\left[
  \begin{array}{ c }
     \bar{T}_\mathbf{k} \\
     \bar{\mathbf{X}}_\mathbf{k}
  \end{array} \right]e^{i(\mathbf{k}\cdot\mathbf{x}-\omega t)}. 
\end{equation}
In Section 2, we met a similar plane wave when we studied the matter field. For our current applications, instead of applying Lorentz covariant plane waves to describe the physical oscillations of matter, we will use them to characterize the fluctuations in spacetime geometry caused by the proper time oscillation. The plane wave defined in Eq. (\ref{eq:array3}) describes spacetime geometrical effects and not physical oscillations of matter.

We can apply $\bar{\xi}_{t\mathbf{k}}$ from Eq. (\ref{eq:array3}) to carry out the decomposition for the proper time oscillation at $\mathbf{x}_0$. However, $\bar{\xi}_{t\mathbf{k}}$ is only the 0-component of a Lorentz covariant plane wave; the spatial component $\bar{\xi}_{\mathbf{x}\mathbf{k}}$ cannot be neglected. Thus, if we superpose the plane waves $\bar{\xi}_{t\mathbf{k}}$ to obtain the proper time oscillation at $\mathbf{x}_0$, there will have spatial oscillations associated with the superposition of $\bar{\xi}_{\mathbf{x}\mathbf{k}}$. These spatial oscillations are essential in our relativistic theory. 

In spherical coordinates, the proper time oscillation and the radial oscillations revealed after the superposition are summarized as follows \cite{Yau20a}:

At $r=0$,
\begin{equation} \label{eq:r=0t}
\bar{t}_f(t,0)=t-\frac{\sin (\omega_0 t)}{\omega_0},
\end{equation}
\begin{equation} \label{eq:r=0r}
\bar{r}_f(t,0)=0.
\end{equation}

At $r=\epsilon/2 \rightarrow 0$,
\begin{equation} \label{eq:r=ep/2t}
\bar{t}_f(t,\epsilon/2)=t,
\end{equation}
\begin{equation} \label{eq:r=ep/2r}
\bar{r}_f(t,\epsilon/2)=\epsilon/2+\Re_\infty \cos(\omega_0 t),
\end{equation}
where $\Re_\infty$ is the amplitude of radial oscillations, and its magnitude is approaching infinity. Based on our convention adopted, $\bar{t}_f(t,r)$ and $\bar{r}_f(t,r)$ are the time and spatial position observed in the spacetime geometry displaced from the equilibrium state $(t,r)$. Outside the sphere with $r=\epsilon/2$, the spacetime is a vacuum with no oscillations.

Eq. (\ref{eq:r=0t}) describes the same proper time oscillation defined in Eq. (\ref{eq:propparticlet1}). This proper time oscillation is stationary at the spatial origin, as demonstrated in Eq. (\ref{eq:r=0r}). The radial oscillations from Eqs. (\ref{eq:r=ep/2r}) are the results of superposing the spatial component of the Lorentz covariant plane waves. These radial oscillations oscillate about a thin shell $\Sigma_0$ with infinitesimal radius ($r=\epsilon/2 \rightarrow 0$) centered at the origin. As we shall note, the amplitude of the radial oscillation has a magnitude approaching infinity ($\Re_\infty\rightarrow\infty$), which implies the instantaneous radial velocity is also approaching infinity. This result will violate the principles of relativity if the oscillations involve motions of matter. Therefore, we cannot study the radial oscillations as motions that can carry matter through space. Instead, we shall consider the radial oscillation as a spacetime geometrical effect acting on an observer stationary on the thin shell $\Sigma_0$.  

In Minkowski spacetime, a clock stationary anywhere in the coordinate system can be synchronized with the clock of a stationary observer $O$ at spatial infinity. However, this is not the case for an observer $\breve{O}$ stationary on the thin shell $\Sigma_0$. As shown in  Eq. (\ref{eq:r=ep/2t}), the clock of $O$ is synchronized with the clock of a 'fictitious' observer $\ubar{O}$ that follows the radial oscillation defined in Eq. (\ref{eq:r=ep/2r}); $\ubar{O}$ is a fictitious inertial observer in the oscillating frame. On the other hand, an observer $\breve{O}$ placed on the thin shell will oscillate relative to $\ubar{O}$. Since the clocks of $O$ and $\ubar{O}$ are synchronized, the clocks of $\breve{O}$ and $O$ cannot be synchronized, albeit the two spatially far apart observers are physically stationary relative to one another. These conditions imply the spacetime geometry (or metrics) at $O$ and $\breve{O}$ are different; a result due to the fictitious oscillation's effects on $\breve{O}$. However, before we proceed further, we shall recall that the instantaneous velocity of the fictitious radial oscillations is approaching infinity. To apply our knowledge in relativity, we will first study a thin shell with a finite radius that has an instantaneous fictitious velocity of less than the speed of light.

In refs. \cite{Yau17,Yau20a}, we investigated a similar timelike hypersurface $\Sigma$ with finite radius $\breve{r}$. On the surface of $\Sigma$, we apply the same fictitious oscillations but with instantaneous velocities $\bar{v}_{f}({t})$ less than the speed of light, i.e.,
\begin{equation} \label{eq:tfrs}
\bar{t}_f(t,\breve{r})=t,
\end{equation}
\begin{equation} \label{eq:rfrs}
\bar{r}_f(t,\breve{r})=\breve{r}+\Re \cos(\omega_0 t),
\end{equation}
\begin{equation} \label{eq:vfrs}
\bar{v}_{f}({t},\breve{r})=\frac{\partial \bar{r}_{f}({t},\breve{r})}{\partial {t}}=-{\Re} \omega_0 \sin(\omega_0 {t}), 
\end{equation}
where $\Re$ is the amplitude of radial oscillation and ${\Re} \omega_0<1$. We can apply relativity to analyze the effects on the observer $\breve{O}$ stationary on the thin shell's surface. However, apart from the instantaneous velocity, the fictitious oscillation also has displacement relative to the thin shell $\Sigma$. From what we have learned about the properties of a harmonic oscillator, we expect the total energy generated by the instantaneous velocity and displacement to be constant over time. Therefore, our system has a time translational symmetry, as predicted by Noether's theorem. The total effects of the instantaneous velocity and displacement on $\breve{O}$ are constant over time.

Observer $\breve{O}$ on the thin shell is stationary relative to observer $O$ at spatial infinity. We can express the infinitesimal coordinate increments $(dt,dr)$ of two events observed by $O$ in terms of the infinitesimal coordinate increments $(d\breve{t},d\breve{r})$, for the same two events observed by $\breve{O}$,
\begin{equation} \label{eq:array1}
\left[
  \begin{array}{ c }
     dt \\
     dr
  \end{array} \right]
=     
\left[
  \begin{array}{ c c }
     {\Upsilon^t}_{\breve{t}} & 0 \\
     0 & {\Upsilon^r}_{\breve{r}}
  \end{array} \right]
\left[
  \begin{array}{ c }
     d\breve{t} \\
     d\breve{r}
  \end{array} \right]. 
\end{equation}
The two off-diagonal terms of the transformation matrix $\Upsilon$ are zeros, which are deduced from the followings: 1) The basis vectors of $O$ and $\breve{O}$ are parallel for two observers stationary relative to one another, i.e., 
$\vec{e}_{\breve{t}} \parallel \vec{e}_{t}$ and $\vec{e}_{\breve{r}} \parallel \vec{e}_{r}$. 2) The basis vectors in the temporal and spatial directions are orthogonal in the local frames of $O$ and $\breve{O}$, i,e., $\vec{e}_{t} \cdot \vec{e}_{r}=0$ and $\vec{e}_{\breve{t}} \cdot \vec{e}_{\breve{r}}=0$. Under the two conditions, we have
\begin{equation}
\Upsilon^t_{\breve{r}}=\vec{e}_{t} \cdot \vec{e}_{\breve{r}}=0,\quad \text{and} \quad \Upsilon^r_{\breve{t}}=\vec{e}_{r} \cdot \vec{e}_{\breve{t}}=0.
\end{equation}

At $t=t_m=\pi/(2\omega_0)$, the fictitious displacement $\bar{r}_d$(=$\bar{r}_f-\breve{r}$) from Eq. (\ref{eq:rfrs}) is zero, but the instantaneous velocity from Eq. (\ref{eq:vfrs}) is, 
\begin{equation} \label{eq:v+fmax}
\bar{v}_{f}(t_m,\breve{r})=\bar{v}_{fm}=-{\Re} \omega_0.
\end{equation}
Therefore, observer $\breve{O}$ on the thin shell is traveling at a velocity $\ubar{v}_{fm} =-\bar{v}_{fm}(={\Re} \omega_0<1)$ relative to the fictitious inertial observer $\ubar{O}$ without displacement from the equilibrium. We can apply relativity to study the properties of a moving observer, albeit the motion is in the fictitious frame. At this instant, the measurements by $\breve{O}$ will undergo length contraction and time dilation relative to the fictitious observer $\ubar{O}$. However, as we shall recall, $\ubar{O}$ is a fictitious inertial observer with its clock synchronized with $O$ at spatial infinity. Although $\breve{O}$ remains stationary with $O$, its measurements will undergo the same length contraction and time dilation relative to $O$. Based on these arguments, we can write the two diagonal terms of the transformation matrix $\Upsilon$ as,
\begin{equation} \label{Upsilon00}
{\Upsilon^t}_{\breve{t}}=[1-(\bar{v}_{fm})^2]^{-1/2}=(1-{\Re}^2 \omega_0^2)^{-1/2},
\end{equation}
\begin{equation} \label{Upsilonrr}
{\Upsilon^r}_{\breve{r}}=[1-(\bar{v}_{fm})^2]^{1/2}=(1-{\Re}^2 \omega_0^2)^{1/2}.
\end{equation}
 Since the system has a time translational symmetry, we can extend these results to all other times. Based on Eqs. (\ref{eq:array1}), (\ref{Upsilon00}), (\ref{Upsilonrr}), and the time translational symmetry, the line element on the thin shell $\Sigma$ is a constant over time, i.e.,
\begin{equation} \label{eq:metric r=rs}
ds^2=[1-{\Re}^2 \omega_0^2]dt^2-[1-{\Re}^2 \omega_0^2]^{-1}dr^2-\breve{r}^2d\Omega^2.
\end{equation}

Following the concepts developed in relativity for a spherically symmetric system, the line element of the external spacetime outside this thin shell $\Sigma$ is a Schwarzschild field,
\begin{equation} \label{eq:metric}
ds^2=[1-\frac{\breve{r} {\Re}^2\omega_0^2}{r}]dt^2-[1-\frac{\breve{r} {\Re}^2\omega_0^2}{r}]^{-1}dr^2-r^2 d\Omega^2.
\end{equation}
In defining this line element, we have adopted the Schwarzschild coordinate system. Introducing a mass for the thin shell, i.e.,
\begin{equation} \label{eq:m}
m=\frac{\breve{r} {\Re}^2\omega_0^2}{2},
\end{equation}
and substitute $m$ into Eq. (\ref{eq:metric}), we obtain the line element of the Schwarzschild spacetime outside a massive thin shell,
\begin{equation} \label{eq:Schwarzschildmetric}
ds^2=[1-\frac{2m}{r}]dt^2-[1-\frac{2m}{r}]^{-1}dr^2-r^2d\Omega^2.
\end{equation}

By Birkhoff's theorem \cite{Birkhoff23,Schmidt13}, the thin shell $\Sigma$ can be contracted while the external geometry remains Schwarzschild as long as the equivalent mass $m$ from Eq. (\ref{eq:m}) remains constant. As our thin shell $\Sigma$ is contracted to $\breve{r}=2m$, it will meet a coordinate singularity when the instantaneous velocity reaches the speed of light. However, the amplitude,
\begin{equation} \label{eq:R}
{\Re}=\sqrt{\frac{2M}{\breve{r}\omega_0^2}},
\end{equation} 
and the metrics developed are well-defined until they reach $\breve{r}=0$. Although the instantaneous velocity exceeds the speed of light, there is no violation of relativity. The fictitious oscillations are spacetime geometrical effects with no transportation of matter through space. Therefore, the thin shell $\Sigma$ can be contracted to an infinitesimal radius with $\Re \rightarrow \infty$, the same thin shell $\Sigma_0$ generated around the proper time oscillator as described in Eqs. (\ref{eq:r=ep/2t}) and (\ref{eq:r=ep/2r}).  These results confirm that the spacetime outside a stationary proper time oscillator is Schwarzschild.

The external spacetime outside a proper time oscillator is curved. Here, we show that proper time oscillation can set up a direct link between matter and spacetime. The theory paints a simple picture: 'The proper time oscillator exerts fictitious radial oscillations on a thin shell with an infinitesimal radius. These radial oscillations alter the spacetime metric on the thin shell's surface and curve the surrounding external spacetime. In turn, the curved spacetime tells other matter how to react in the presence of the proper time oscillator' \cite{Yau20a}.

\section{Review of a Proper Time Oscillator}

The following subsections elaborate the properties of a proper time oscillator. We will clarify certain concepts of the theory.

\subsection{Proper Time Oscillator vs. Standard Theories}

Let us compare the properties derived from our theory against those predicted by quantum theory and general relativity. Regarding general relativity, we have demonstrated that the gravitational field of a proper time oscillator, when assumed stationary in space, is a Schwarzschild field, the same as predicted by general relativity for a rest point mass. The proposed theory can explain how the intrinsic structures of matter influence the spacetime geometry, establishing a more direct correlation between spacetime and matter. However, we should remember that the quantum effects were neglected when we developed the gravitational field. The proper time oscillator was assumed to be a classical object in the analysis. A complete quantum gravity theory must be developed to understand the gravitational properties of a proper time oscillator fully.

Quantum theory is one of the most quantitatively accurate theories in science. Numerous experiments have tested the theory. We will demonstrate below that the properties of a proper time oscillator stay consistent with the predictions of quantum theory for a bosonic field until we reach a very high energy level, where the oscillations of matter in time and space become significant.

As shown in Section \ref{sec:3.1}, a $\zeta$ field contains information for the oscillations of matter in the system. Applying Eqs. (\ref{eq:zetat1}) and (\ref{eq:zetax1}), the temporal and spatial oscillation displacements, $\zeta_t$ and $\zeta_\mathbf{x}$, are the derivatives of $\zeta$ in respect to $t$ and $\mathbf{x}$ respectively. Also, as defined in Eq. (\ref{eq:varphi2}), a real scalar field $\varphi$ can be obtained from $\zeta$, where $\varphi$ satisfies the Klein-Gordon equation. Apart from the proper time oscillations, the real scalar field $\varphi$, derived from $\zeta$, has the exact properties of a bosonic field in quantum field theory. 

Applying $\zeta$ (or $\varphi$) to describe the properties of a real scalar field and its oscillations is analogous to using a four-potential $A^\mu$ in electromagnetic theory. The electric and magnetic fields, $E$ and $\mathbf{B}$, are components of the electromagnetic tensor, which is the exterior derivative of $A^\mu$. The properties of an electromagnetic field can be derived from the four-potential $A^\mu$ in lieu of explicitly in terms of the $E$ and $\mathbf{B}$ fields. Similarly, instead of explicitly using the temporal and spatial displacements, $\zeta_t$ and $\zeta_\mathbf{x}$, it is sufficient to apply $\zeta$ (or $\varphi$) in our formulations to describe the dynamics of a field with oscillations of matter in time and space. From $\zeta$, we can obtain $\zeta_t$ and $\zeta_\mathbf{x}$.

The difference between our proposed theory and quantum theory is the additional oscillations of matter in proper time. Our theory makes the exact predictions as the quantum field theory for a bosonic field until the temporal and spatial oscillations become significant. Suppose an experiment is not sensitive enough to detect the effects of the proper time oscillations. In that case, there are no deviations of the predictions between our proposed theory and quantum theory. On the other hand, the oscillations are small and cannot be detected by experiments yet, as we will demonstrate in the following subsection.

\subsection{High Energy Magnification} \label{sec 5.2}

As shown in Eq. (\ref{eq:propparticlet1}), the internal time of a particle evolves with oscillation, implying that the decay of an unstable particle has oscillating rates. Let us examine the magnitude of the proper time oscillations for $\pi^\pm$ ($\omega_0=2.1\times10^{23 }s^{-1}$) and $\pi^0$ ($\omega_0=2.0\times10^{23 }s^{-1}$), which are the lightest spin-0 mesons. From Eq. (\ref{eq:T0}), the amplitudes of the proper time oscillation for $\pi^\pm$ and $\pi^0$ are $4.8\times10^{-24 }s$ and $5.0\times10^{-24 }s$ respectively. Compared with the decay mean lifetime measured for the $\pi^\pm$ of $(2.6033 \pm 0.0005)\times 10^{-8} s$ and $\pi^0$ of $(8.5 \pm 0.2) \times 10^{-17} s$ \cite{Patrignani16}, the proper time oscillations are small and beyond the resolutions that our experiments on particle decays can detect yet.

Next, let us examine the magnitude of oscillations for a relativistic particle. We will consider a normalized plane wave\footnote{Without loss of generality, we have omitted $\mathbf{k}$ when labeling the normalized plane wave in this section.},
\begin{equation} \label{eq:zetan}
\tilde\zeta=\frac{e^{i(\mathbf{k}\cdot\mathbf{x}-\omega t)}}{\sqrt{\omega\omega_0^3}}.
\end{equation}
 From Eq. (\ref{eq:Hamiltonian}), its Hamiltonian density is,
\begin{equation}
\tilde{\mathcal{H}}=\omega/V, 
\end{equation} 
consisting one particle of energy $\omega$ in our system with volume $V$. In this normalized plane wave, the observed particle travels at an average velocity $\mathbf{v}=\mathbf{k}/\omega$.  

As the particle propagates, it oscillates in time and space. From Eqs. (\ref{eq:oscillations}), (\ref{eq:zetat1}) and (\ref{eq:zetax1}), the oscillation displacements are,
\begin{equation} \label{eq:zetant2}
\tilde t_d=\textrm{Re}(\partial_0 \tilde{\zeta})= \mathring{T} \sin(\mathbf{k}\cdot\mathbf{x}-\omega t),
\end{equation}
\begin{equation} \label{eq:zetanx2}
\tilde{\mathbf{x}}_{d}=-\textrm{Re}(\mathbf{\nabla}\tilde{\zeta})=\mathring{\mathbf{X}}\sin(\mathbf{k}\cdot\mathbf{x}-\omega t),
\end{equation}
\begin{equation} \label{eq:Tn}
\mathring{T}=\sqrt{\frac {\omega }{\omega_0^3}},\quad \mathring{\mathbf{X}}=\frac{\mathbf{k}}{\sqrt{\omega_0^3\omega}}.
\end{equation}
As we shall recall, we have adopted a convention similar to the Lagrangian formulation in wave mechanics. A particle propagates with a temporal oscillation amplitude $\mathring{T}$ and spatial oscillation amplitude $\mathring{\mathbf{X}}$.

The oscillations can be magnified by projecting the particle to a higher energy level, e.g., experiments on a particle's arrival time. Let us consider the oscillation amplitudes for a $\pi^+$. As shown in Table 1, the amplitudes are amplified when the particle's speed is increased. At higher energy, the effects of the particle's oscillations will be easier to detect. However, even for a $\pi^+$ particle with an energy 1 TeV, detecting the oscillations is still beyond the reach of our experiments.  

\begin{table}[h]
\begin{center}
\begin{minipage}{174pt}
\caption{Oscillation amplitudes of a $\pi^+$ with different projected energies.}\label{tab1}
\begin{tabular}{@{}lll@{}}
\toprule
$E (GeV)$& $\mathring{T} (s)$ & $\mathring{X} (m)$ \\
\midrule
1\hphantom{00} & \hphantom{0}$1.3\times 10^{-23} $& \hphantom{0}$3.5\times 10^{-15} $  \\
10\hphantom{00} & \hphantom{0}$4.0\times10^{-23}$ & \hphantom{0}$1.2\times 10^{-14} $  \\
100\hphantom{0} &\hphantom{0} $1.3\times10^{-22}$ & \hphantom{0}$3.8\times 10^{-14} $ \\
1000 \hphantom{0} & \hphantom{0} $4.0\times10^{-22}$ & \hphantom{0} $1.2\times 10^{-13} $ \\
\end{tabular}
\end{minipage}
\end{center}
\end{table}

As shown in Eq. (\ref{eq:Tn}), the oscillation amplitudes decrease with increasing particle mass. A heavier particle has smaller amplitudes than a lighter particle if both are projected to the same energy. Since $\pi^+$ is already one of the lightest spin-0 massive bosons and its oscillations at $1TeV$ are not yet detectable, the oscillations of all other massive bosons are also too small for detection. Our theory makes the exact predictions as the quantum field theory for a bosonic field until the effects of a particle's oscillations become significant. However, those oscillations have not yet reached a level that are detectable.

So far, our discussions have been limited to bosons. What about fermions? If a boson's mass-energy is generated by proper time oscillation, we expect the same can be true for a fermion. As an intrinsic property of a particle, the properties of mass are the same for all massive particles regardless of their spins. In the following discussions, we make a presumption that a spin-1/2 particle also has proper time oscillation. Whether this presumption is valid requires further examinations, which will be delayed to a future paper. Here, we aim to examine the magnitude of oscillations for fermions, assuming they also have proper time oscillations.

Let us consider an electron ($\omega_0=7.6\times10^{20 }s^{-1}$), the lightest elementary particle apart from neutrinos. From Eq. (\ref{eq:T0}), the amplitude of the proper time oscillation is $\mathring{T}_0=1.3\times10^{-21 }s$. Projected at an energy of 1 TeV, the amplitudes from Eq.(\ref{eq:Tn}) are  $\mathring{T}=1.8\times10^{-18 }s$ and $\mathring{\mathbf{X}}=5.6\times10^{-10 }m$. Again, the oscillations are small for detection. 

Next, let us consider a neutrino. As of today, the masses of the three neutrinos are unknown. However, active research has been carried out in this area, with expectations that new physics could be revealed. Because of their extremely lightweight, neutrinos have much larger temporal and spatial oscillations than other particles.

To demonstrate the magnitude of the oscillations, we will assume a neutrino mass of $m=2eV$ \cite{Aseev12,Olive14}.
As shown in Table 2, the amplitudes are much larger than those for an electron with the same energy. Despite a neutrino is hard to detect, and its mass is unknown, the spatial amplitudes projected at high energy could be in the macroscopic scale, e.g., $\mathring{\mathbf{X}}= 7.0$cm at $E=$1Tev. Because they are extremely lightweight, neutrinos can be projected at very high speed, amplifying the oscillations for possible measurements. If our assumptions made for a neutrino are correct, experiments on neutrinos could provide some hints on how to test our theory. Further investigations are required.

\begin{table}[h]
\begin{center}
\begin{minipage}{174pt}
\caption{Oscillation amplitudes of a neutrino with different projected energies with assumed mass $m=2eV$.}\label{tab1}
\begin{tabular}{@{}lll@{}}
\toprule
$E (GeV)$& $\mathring{T} (s)$ & $\mathring{X} (cm)$ \\
\midrule
1\hphantom{00} & \hphantom{0}$7.4\times 10^{-12} $& \hphantom{0}$0.22$ \\
10\hphantom{00} & \hphantom{0}$2.3\times10^{-11}$ & \hphantom{0}$0.70$ \\
100\hphantom{0} &\hphantom{0} $7.4\times10^{-11}$ & \hphantom{0} $2.2$ \\
1000 \hphantom{0} & \hphantom{0} $2.3\times10^{-10}$ & \hphantom{0} $7.0$ \\
\end{tabular}
\end{minipage}
\end{center}
\end{table}

\subsection{Forward Flow of Time}

The proper time oscillation is an assumed intrinsic property of a particle. A particle's varying internal time rate is not a product of the time dilation in a moving frame or gravitational field. As an oscillator, the internal time of a particle never travels backward in time. From Eq. (\ref{eq:propparticlet1}), the internal time rate relative to the coordinate time is bounded between 0 and 2; its average rate is 1. Therefore, a particle's internal time flows only in the forward direction.  

\begin{figure}[htp]
    \centering
    \includegraphics[width=10cm]{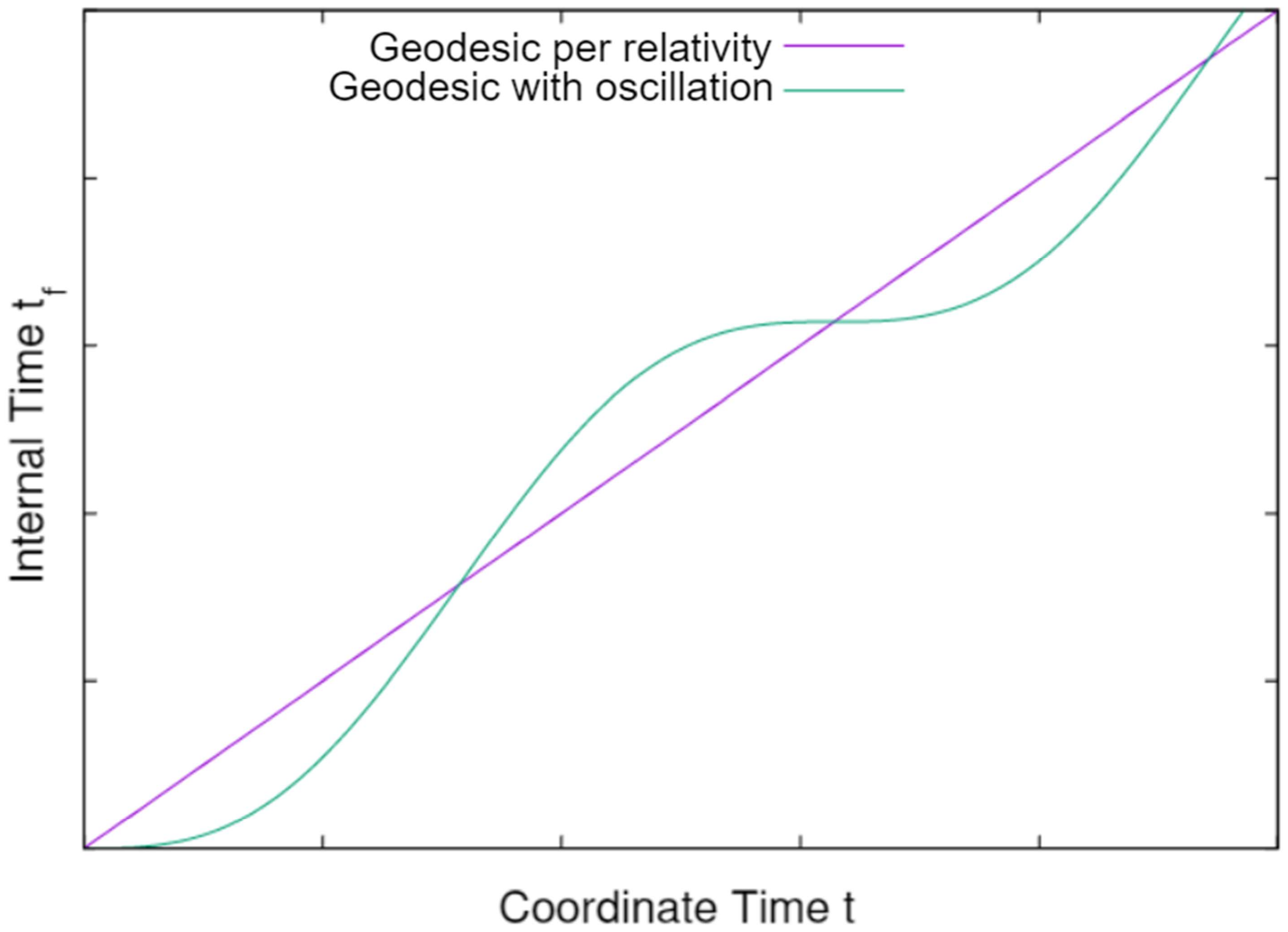}
        \label{fig:Oscillation}
\caption{Proper time oscillation of a $\pi^0$ particle with $\omega_0=2.0\times 10^{23}s^{-1}$ and $T_0=5.0\times 10^{-24}s$. The internal time never travels back to the past.}
\end{figure}

Fig. 1 depicts the nature of the proper time oscillation for a $\pi^0$ particle. The purple line is the 'smooth' time-like geodesic according to relativity. The green line is the internal time of the particle with oscillation obtained from Eq. (\ref{eq:propparticlet1}). The cycle is repeated, but there is not a single moment that the internal time reverts to its past. A particle with proper time oscillation does not travel backward in time.

In a Minkowski spacetime, the internal time of a particle evolves tightly with the coordinate time. In Section \ref{sec 5.2}, we demonstrate that the proper time oscillation amplitude, and thus, the deviation of the particle's internal time from the coordinate time, is very small. The oscillating internal time of a particle can never make a wild jump to the far future. 


\subsection{Time and Space Symmetry} \label{sec5.3}

General relativity treats time and space on an equal footing, which differs from how time is treated in quantum theory. Time enters quantum theory as a parameter and not as an operator, contrasting the position $\mathbf{x}$ of a particle, which is the eigenvalue of an operator. Unsurprisingly, the asymmetrical way of treating time and space has created a constellation of problems when we try to reconcile quantum theory with general relativity, e.g., the problem of time.

Attempts to promote time to an operator have met many difficulties. One of the reasons why time is not treated as a self-adjoint operator can be traced back to Pauli. According to Pauli \cite{Pauli80,Srinivas81}, a time operator and the Hamiltonian of a system should satisfy a commutation relation, $[H,t] =-i$. As a universal time operator, its spectrum should continuously span the entire real line. On the other hand, the Hamiltonian $H$ is bounded from below for a typical physical system. If the Hamiltonian $H$ forms a conjugate pair with the time operator $t$, its spectrum shall also be continuous and unbounded. However, this conclusion would contradict our observation that the energy for a physical system is typically bounded. Based on these reasonings, time is generally not treated as an operator in the standard formulations of quantum theory.

Another issue is that a time operator is surprisingly complicated to develop. In a relativistic theory, we can promote the coordinate time $t$ as an operator. In that case, a particle's proper time $\tau$ can be taken as the time parameter, allowing us to define operators $X^\mu (\tau)$ with $X^0=t$. However, this line of thinking has encountered a problem of 'many times', resulting in an infinite redundancy of descriptions, which turns out to be very complicated to account for. Because of this, another approach has been adopted to put time and space on an equal footing. The position is demoted from an operator to match the status of time as a label in the standard quantum field theory.



In this paper, we have considered a symmetry between time and space in a matter field. However, the symmetry we have investigated is for harmonic oscillations and not coordinate time. 
As a proper time oscillator, the internal time $t_f$ of a particle propagates with oscillation about the coordinate time $t$, i.e.,
\begin{equation}
t_d=t_f-t.    
\end{equation}
When considering the temporal displacement $t_d$, the coordinate time $t$ is the equilibrium point of the oscillation. This temporal oscillation is an analogy of the oscillation in space. A difference is that the equilibrium point of a spatial oscillator is stationary in space. 

As shown in Section \ref{sec3.3}, the temporal oscillation displacement can be treated as a self-adjoint operator analogous to its spatial counterpart. However, coordinate time is a parameter and not an operator, the same as what is adopted in quantum theory. The symmetry we have considered here is between the temporal and spatial oscillations. 

Coordinate time remains a parameter in our theory. Whether there is a more profound symmetry between space and time in quantum theory, where coordinate time can be treated as an operator, is beyond the scope of our investigations. However, we have arrived at some interesting results by adopting the space and time symmetry for harmonic oscillations. As demonstrated, we can reconcile the properties of a bosonic field and the gravitational field of a point mass by treating a particle as a proper time oscillator. 

The introduction of the internal time and temporal oscillation operators in Section \ref{Sect3.4} have no conflict with Pauli's theorem. The commutation relations established in Eqs. (\ref{zeta-zetat}) and (\ref{zeta-zetat2}) do not involve energy. The internal time and energy do not form a conjugate pair. Therefore, there is no restriction on the internal time spectrum, albeit the system's Hamiltonian is bounded from below. The internal time and temporal oscillation can be treated as self-adjoint operators without contradicting Pauli's theorem.

\section{Conclusions and Discussions}
\label{sec:6}

In quantum field theory, particles are treated as sets of coupled oscillators with their own de Broglie's internal clock. Interestingly, the proper time oscillators meet these criteria. In this paper, we demonstrate that the properties of a zero-spin bosonic field and Schwarzschild spacetime can be reconciled from an assumption that matter oscillates in proper time. The properties obtained from our assumption, e.g., self-adjoint internal time operator, proper time uncertainty relation, and generation of a gravitational field, could reduce some of the differences between quantum theory and general relativity. 

We have also clarified a few concepts of the theory: 1) The time and space symmetry considered is for harmonic oscillations, not coordinate time. In our formulations for the quantum theory, coordinate time remains a parameter, the same as what is adopted in quantum field theory. 2) A particle with proper time oscillation can never travel back to its past. The internal time of a particle flows only forward. 3) Our theory makes the exact predictions as the quantum field theory for a bosonic field until a particle's proper time oscillation effects become significant. However, those oscillations have yet to reach a detectable level. 4) The oscillations in time and space can be magnified by projecting a particle to higher energy. Because a neutrino is extremely lightweight, the particle can be projected at a very high speed. If our assumptions made for a neutrino are correct, experiments on neutrinos could provide some hints on how to test our theory.

\end{document}